\documentclass[journal=MAMOBX,manuscript=article]{achemso}

\usepackage[version=3]{mhchem} 
\usepackage{graphicx}
\usepackage{bm}
\usepackage{amsmath}
\usepackage{amssymb}
\usepackage{graphicx}
\usepackage{xcolor}
\usepackage{makecell}
\usepackage[version=3]{mhchem} 
\usepackage{multirow}
\usepackage{ulem}

\newcommand{\etal}{\textit{et al}. }

\newcommand{\del}[1]{\unskip}
\author{Le Qiao}
\email{le.qiao@uni-mainz.de}
\affiliation{Institut für Physik, Johannes Gutenberg-Universität Mainz, D55099 Mainz, Germany}%
\author{Daniel A. Vega}
 \affiliation{Instituto de Física del Sur (IFISUR), Consejo Nacional de Investigaciones Científicas y Técnicas (CONICET), Universidad Nacional del Sur, 8000 Bahía Blanca, Argentina}%
\author{Friederike Schmid}%
 \email{friederike.schmid@uni-mainz.de}
\affiliation{Institut für Physik, Johannes Gutenberg-Universität Mainz, D55099 Mainz, Germany}%

\title{Stability and Elasticity of Ultrathin\\ Sphere-Patterned Block
Copolymer Films}

\begin{document}


\begin{abstract}
Sphere-patterned ultrathin block copolymers films are potentially
interesting for a variety of applications in nanotechnology. We use
self-consistent field theory to investigate the elastic response of
sphere monolayer films with respect to in-plane shear, in-plane
extension and compression deformations, and with respect to bending.
The relations between the in-plane elastic moduli is roughly
compatible with the expectations for two-dimensional elastic systems
with hexagonal symmetry, with one notable exception: The pure shear and the simple shear
moduli differ from each other by roughly $20 \%$. Even more importantly, the bending constants are
found to be negative, indicating that free-standing block 
copolymer membranes made of only sphere  monolayer are inherently unstable
above the glass transition. Our results are discussed in view of experimental
findings.
\end{abstract}

\section{Introduction}
The self-organization of block copolymers has captured the interest of
researchers for many decades, not only due to the richness of emerging
structures, but also because it provides a cheap and efficient route
to nanofabricating novel materials with tunable properties
\cite{yangBlockCopolymerNanopatterning2022,
karayianniBlockCopolymerSolution2021}.
In particular, self-assembling block copolymer thin films are of
high technological interest for a variety of applications such as
antireflective coatings, nanoscale templates, metal nanodots,
conducting membranes, and organic optoelectronics, since they form
patterns on length scales that are not easily accessible to
traditional lithographic techniques \cite{kulkarniThinFilmBlock2022,
huangBlockCopolymerThin2021}.

The majority of the applications of block copolymer (BCP) films require
the microstructure to be almost perfectly ordered on a large scale
\cite{albertSelfassemblyBlockCopolymer2010, huangBlockCopolymerThin2021}.
Spontaneous self-assembly usually proceeds via the growth of ordered
domains from several nucleation centers at random positions with random
orientations. This results in multi-domain structures with numerous
topological defects at the interfaces, which they lack long-range order
\cite{liDefectsSelfAssemblyBlock2015b,albertSelfassemblyBlockCopolymer2010,
garciaDefectFormationCoarsening2015}. Various methods, including
electric field techniques \cite{morkvedLocalControlMicrodomain1996},
mechanical shear alignment
\cite{marencicOrientationalOrderSphereForming2007,Kwon2014}, zone-annealing
\cite{nowakPhotothermallyDirectedAssembly2020}, and geometric curvature
constraints \cite{vuCurvatureGuidingField2018}, have been proposed for
controlling morphology formation and local alignment. The central idea
of these methods is to induce a phase transition or reorientation by
applying external fields that cause a deformation in the BCP film, thus
altering its free energy. The response of thin films to strain can be
quantified by elastic constants, which, in turn, depend on the type of
surface pattern.  Understanding the elasticity of microphase separated
BCP films with given target patterns is therefore an essential
ingredient for evaluating the potential of alignment strategies and
of options to control their self-assembly and eliminate defects on larger
scales.

In the past few decades, there has been significant research on the
elastic behavior of microphase-separated bulk BCP melts in different
phases such as cylinder, lamellar, sphere, and gyroid
\cite{kossuthViscoelasticBehaviorCubic1999,
tylerLinearElasticityCubic2003,
thompsonElasticModuliMultiblock2004,liElasticPropertiesLine2013}. Thin films are much less well understood. Both in the bulk and in thin films,
the elastic properties and the micro-domain morphology of BCP melts
depend on the BCP architecture, the volume fractions of the blocks, and
the strength of monomer interactions. In thin films, the film thickness
and the surface energy are additional important factors that
influence their microstructure
\cite{batesBlockCopolymersDesigner1999,
saitoFilmThicknessStrain2022}. For example, the strength and
selectivity of surface interactions determines their overall wetting
behavior as well as island and hole formation due to incommensurability
effects \cite{albertSelfassemblyBlockCopolymer2010}.
The elastic behavior of BCP films can also be expected to be
quite different from that in the bulk.  In particular, monolayers of
sphere-forming thin films consist of arrays of self-assembled domains
with hexagonal symmetry\cite{albertSelfassemblyBlockCopolymer2010,
marencicShearAlignmentRealignment2010} (see Fig.~\ref{fig:system_scheme}a
), which differs from the dominant
body-centered cubic order in the bulk. 
This has consequences for the success of alignment strategies.
Angelescu \etal found that in-plane shear can effectively enhance
the alignment for BCP films containing two layers of spheres, however,
they did not succeed to align monolayer films.
\cite{angelescuShearInducedAlignmentThin2005} On the other hand,
(in-plane) bend deformations can be enforced much more easily in thin
films than in the bulk, which provides additional opportunities for
alignment strategies \cite{vuCurvatureGuidingField2018}.

This motivates the present work. Our goal is to investigate the elastic response of thin symmetric sphere-patterned BCP films to various types of elastic deformations. Directly studying the elastic behavior of such ultra-thin films presents significant challenges \cite{liElasticPropertiesLine2013}. Self-consistent field theory (SCFT) has emerged as a powerful tool for describing and predicting equilibrium BCP structures \cite{matsenFastAccurateSCFT2009, tylerLinearElasticityCubic2003}. Within the realm of thin films, SCFT has been extensively employed to explore the correlation between phase behavior and film thickness \cite{geisinger99, hurSCFTSimulationsThin2009, mishraSCFTSimulationsOrder2011, liPhaseDiagramDiblock2013}. In our study, we first utilize SCF calculations to establish a stable sphere-patterned single-layered film structure. Subsequently, we match the commensurability of the SCF-calculated film with the sphere-forming PS–PHMA diblock copolymer we obtained via spin-coating. Finally, we analyze the impact of mechanical strain on the stability and elasticity of a monolayer BCP film in the hexagonal phase, with the aim of elucidating why ultra-sphere-patterned thin films exhibit reduced stability compared to other structures.

\section{Methods}
\begin{figure}[htbp]
\begin{center}
\includegraphics[scale=1]{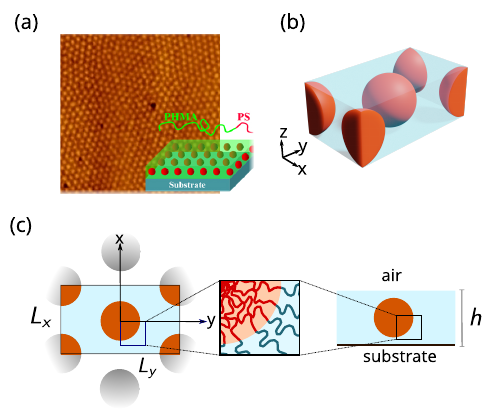}
\end{center}
\caption{(a) AFM phase image showing a monolayer of sphere-forming
PS-PHMA diblock copolymer deposited onto a flat bare silicon wafer.
The image size is 1.0 $\mu$m $\times$ 1.0 $\mu$m. The inset shows a
schematic of a monolayer of spheres on top of a flat substrate. (b)
The SCFT calculations are conducted in a periodic cuboid cell with
in-plane periodic boundaries (x-y plane) and Dirichlet boundaries in
the z direction. (c) 2D schematics showing the middle slices of the
calculation cell in the xy plane and the yz plane.}
\label{fig:system_scheme}
\end{figure}

\textbf{Experiment.} The PS-PHMA diblock employed here was synthesized via sequential living anionic polymerization of styrene and n-hexyl methacrylate, in tetrahydrofuran at -78$^{\circ}$C, using 10 eq of LiCl to s-butyllithium initiator. The sphere-forming diblock had block molar masses of 18 and 95 kg/mol (see refs. \cite{varshney1990anionic,Kwon2014,Garcia2014} for more details ). 
A monolayer of diblock copolymer was made by spin-coating a solution of PS-PHMA in toluene ($\sim$ 1 wt. $\%$) onto silicon wafer (SiliconQuest, with native oxide).  The thickness of the film was controlled through the spin speed. The silicon wafer was pre-washed at least five times with toluene and dried under flowing nitrogen prior to use.  The polymer film thickness was measured using a PHE101 ellipsometer (Angstrom Advanced Inc.; wavelength = 632.8 nm). The morphology was characterized at room temperature using atomic force microscopy (Innova, Bruker) operated in tapping mode using uncoated silicon tips having a cantilever length of 125 $\mu$ m, spring constant of 40 N/m, and resonant frequency of 60 $-$ 90 kHz, purchased from NanoWorld. Since the PS spheres are glassy and the PHMA matrix rubbery at room temperature, AFM phase imaging of the films reveals the underlying structure and the local order developed by the block copolymer. To improve the image quality, the micrograph was flattened and filtered by performing a discrete Fourier transform to remove high and low frequency noise. Fig.~\ref{fig:system_scheme}a shows the pattern developed by the arrays of PS spheres after 12 hours of thermal annealing at 125 $^{\circ}$C. Thin films thickness $ \varepsilon \sim$ 35 nm.

\textbf{SCF calculations.} We consider a melt of coarse-grained, flexible AB
diblock copolymer molecules with a degree of polymerization $N$,
symmetric segment length $b$, and a volume fraction $f$ for the A-block.
The interactions between the polymer chains are characterized by the
dimensionless incompatibility parameter $\chi N$ and compressibility parameter
$\kappa N$. We consider symmetric films, corresponding to a situation
where the surface interactions on both sides of the films are
identical, both showing a preference for A-blocks. The SCFT
calculations are conducted in a periodic cuboid cell with a volume of
$L_xL_yh$ (Fig.~\ref{fig:system_scheme}b-c) in the grand canonical
ensemble, i.e., a fixed chemical potential of the copolymer melt is
used.  More details can be found in Supplementary Information S1.

We first obtain the stable sphere structure by minimizing the free
energy per unit area of the melt, $F/A$, with respect to the transverse
periodicity of the sphere domains ($L_x$) and cell thickness ($h$). The
transverse cell shape in this case is fixed to $L_y/L_x=\sqrt{3}$,
corresponding to a hexagonal pattern. We will see below that this
indeed corresponds to the energetically preferred shape. In our
calculation, the chemical potential of the copolymer melt is fixed to
$\mu=(2.25+\ln G)$, where $G=\rho_0 R_g^3/N$ is the dimensionless Ginzburg
parameter, $\rho_0$ is the average monomer density, and
$R_g=b\sqrt{N}/6$ is the radius of gyration of one polymer chain. The
modified diffusion equation is solved in real space using the
Crank-Nicolson scheme and the alternative direction implicit (ADI)
method (numerical details can be found in Supplementary Information S3). We use
a contour discretization of $\Delta s=0.0006$ and a spatial
discretization using a grid system of $N_xN_yN_z=20\times34\times350$. A
finer discretization is chosen in the $z$ direction to minimize the
discretization error arising from the external surface interaction
potential applied at the interface.
\section{Results and discussion}
\begin{figure}[htbp]
\begin{center}
\includegraphics[scale=1]{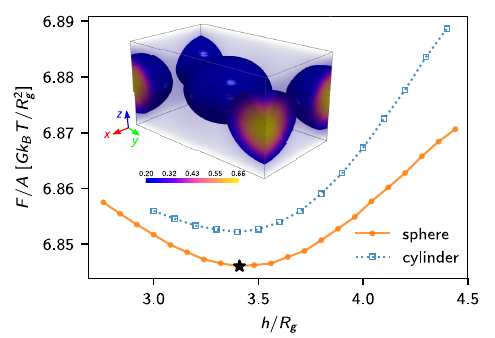}
\end{center}
\caption{Grand canonical free energy per area ($F/A$) as a function of film thickness $h$. The optimum thickness corresponds to the free energy minimum is marked with $\star$. Inset: the SCFT calculated density profile of B-block for $h=3.4~R_g$.} 
\label{fig:1layer_opt}
\end{figure}

Previous SCFT calculations predict that the sphere phase exists in a
very narrow window ($f \in [0.757, 0.790]$ for $\chi N = 20$) in the
bulk phase diagram of AB diblock-copolymers
\cite{matsenFastAccurateSCFT2009}. Here we choose $f = 0.76$, $\chi N =
20$, and $\kappa N = 25$. For this value of the compressibility parameter, the density inside the film is the same for all simulations within 0.03\%.
The strength of surface interaction between
surface and copolymer blocks A and B per surface area is $\gamma_AN =
-6$ and $\gamma_BN = -1$, in units $Gk_BT/R_g^2$.
Fig.~\ref{fig:1layer_opt} shows the SCFT results for the free energy per
area as a function of the film thickness $h$. In an SCFT study of a
similar system, Li et al.\ obtained a transition from sphere (at small
$h$) to cylindrical (at larger $h$) around $h \sim 2.4 R_g$
\cite{liPhaseDiagramDiblock2013}.  Our SCFT calculations, instead,
predict a stable sphere phase for film thickness $h/R_g\in [2.7,4.5]
$. We attribute this difference to the varied asymmetries in the surface interactions with A- and B-blocks, a factor we found to be critical in constructing the phase diagram of thin films. We have tested
that the free energy of the (metastable) cylinder phase (blue dotted curve in
Fig.\ 2) is indeed higher than that of the other phases with our choice
of parameters. The optimal lateral inter-sphere spacing in our system is
found to be $L_x = L_y/\sqrt{3} = 3.7\,R_g$ and the optimum thickness is
$h^* = 3.4\,R_g$, corresponding to an incommensurability parameter of
$h^*/L_x \approx 0.92$.  This is in reasonably good agreement
with our experimental result, $h^*/L_x \approx 0.97$. Matching the
periodicity and the film thickness to the experiments, we can estimate
$R_g \approx 10.5\,nm$, and thus $G \approx 6.2$, by assuming an average
copolymer density of $1\,g/cm^3$ at room temperature
\cite{kimLargeAreaNanosquareArrays2014}.

\begin{figure}[htbp]
\begin{center}
\includegraphics[scale=1.4]{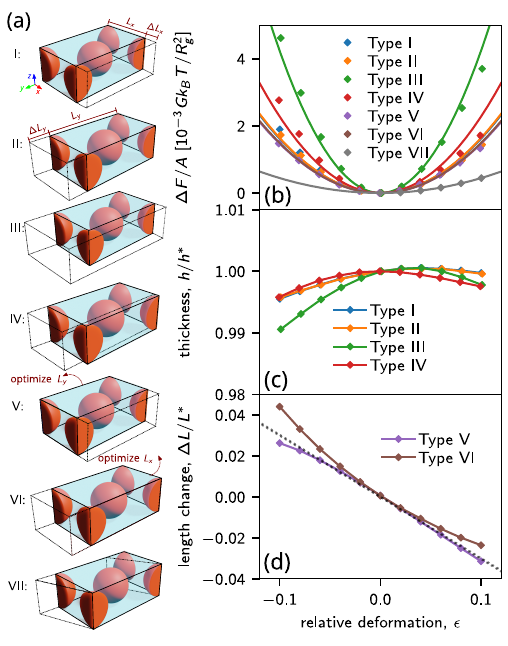}
\end{center}
\caption{(a) Different types of deformations.(b) Free energy change
per area ($\Delta F/A$) in units of $Gk_BT/R_g^2$ versus
relative deformation ($\epsilon$). The curves result from a collective
parabolic fit of the SCFT data for all seven deformations to
$f(\epsilon)=\frac{1}{2}\hat{K}_{i}\epsilon^2$, where $\hat{K}_i$ is a
function of $K_{11}$, $K_{22}$, $K_{12}$, and $K_{66}$ as shown in
Table~\ref{tab:deformation}, and the subscript $i$ indicates different
deformation types. The curves for type I (blue) and type II (orange)
are almost identical. Fitting parameters: $K_{11}=0.32(3\pm7) $, $K_{22}=0.32(8\pm7)$, $K_{12}=0.09(8\pm4) $ and $K_{66}=0.089$ (c) Relative thickness $h/h^*$ versus relative
deformation for Type I-IV. (d) Relative change ($\Delta L/L$) in $L_y$
and $L_x$ for Type V and VI. The slope of the dashed line gives the
Poisson ratio $\nu_x\approx\nu_y\equiv-\frac{\Delta L/L}{\epsilon}\approxeq 0.3$.} 

\label{fig3:stretching_el}
\end{figure}

To calculate the elastic constants, we analyze the free energy response
of the system to small perturbations of the size or shape of the cell
used in the SCFT calulations (Fig.~\ref{fig:system_scheme}b). In the
linear regime, the general form of the resulting free energy change
should take the form $\Delta
F=\frac{1}{2} K_{ijkl} u_{ij} u_{kl}$, where $K_{ijkl}=\frac{\partial^2
F}{\partial u_{ij}\partial u_{kl}}$ is the rank-four elastic modulus
tensor, and $u_{ij}=\frac{1}{2}\left(\frac{\partial u_i}{\partial
x_j}+\frac{\partial u_j}{\partial x_i}\right)$ is the strain tensor
($i,j,k,l=1,2,3$). Since only in-plane extension/compression and shearing
are considered, the elastic free energy for the BCP film can generally be
decomposed as \cite{boal02}
\begin{equation}
\label{eq:energy_2D}
\begin{aligned} 
  \Delta F/A= \frac{1}{2} \left(K_{11}u_{xx}^2
     + K_{22}u_{yy}^2 +2K_{12} u_{xx} u_{yy}
     +4 K_{66} u_{xy}^2\right)
\end{aligned},
\end{equation}
where $K_{11}$, $K_{22}$, and $K_{12}$ are the extensional moduli and
$K_{66}$ is the shear modulus. For simplicity purposes, we use the Voigt notations for the moduli, where the the subscripts 1, 2 and 6 correspond to $xx$, $yy$ and $xy$. We calculate the elastic moduli
by numerically evaluating the variation in the SCFT free energy of
the BCP film after a small deformation, such that the film can still
re-equilibrate. The second derivative of the free energy with respect
to strain gives the elastic moduli. We impose seven different types of
in-plane deformations to the film by varying cell size and shape as
follows as shown in Fig.~\ref{fig3:stretching_el}a. Table~\ref{tab:deformation} summarizes different types
of deformations and their associated strain components and the combined
elastic moduli that contribute to the free energy change. \del{The the individual elastic modulus from the collective parabolic fit in
Fig.~\ref{fig3:stretching_el}b.} The calculated free energy difference,
$\Delta F/A$, after deformation can be fitted with a parabolic function,
$f(\epsilon)=\frac{1}{2}\hat{K}_{i}\epsilon^2$, where $\epsilon$ is
the relative deformation and $\hat{K}_{i}$ is effective modulus for
deformations type $i$. We derive the effective elastic moduli,
$\hat{K_i}$ for each individual deformation from eq.~\ref{eq:energy_2D}.
By fitting the SCFT data in Fig.~\ref{fig3:stretching_el}b for
all seven deformations collectively, we obtain the elastic modulus
$K_{11}=0.32(3\pm7) $, $K_{22}=0.32(8\pm7)$, $K_{12}=0.09(8\pm4) $ and $K_{66}=0.089$. Each deformation also
induces a small change in the film thickness, shown in 
Fig.~\ref{fig3:stretching_el}b. Figure~\ref{fig3:stretching_el}a shows the different types of
deformations and the changes in free energy as a function of relative
deformations.

\begin{table}[htbp]
\centering
\begin{tabular}{ |c||c|c|c| }
 \hline
 Type & \multicolumn{2}{c|}{ strain} & $\hat{K}_i$ \\
 \hline
 I          & $u_{xx}=\epsilon$ \quad & $u_{yy}=u_{xy}=0$ & $K_{11}$  \\
 II         & $u_{yy}=\epsilon$  \quad &$u_{xx}=u_{xy}=0$&  $K_{22}$  \\
 III        & $u_{xx}=u_{yy}=\epsilon$ \quad& $u_{xy}=0$ & $K_{11}+K_{22}+2K_{12}$\\
 IV         & $u_{xx}=\epsilon$, $(1+u_{yy})(1+u_{xx})=1$ \quad& $u_{xy} =0$ & $K_{11}+K_{22}-2K_{12}$  \\
 V & $u_{xx}=\epsilon$, $\partial \Delta F/\partial u_{yy}=0$ \quad& $u_{xy}=0$ & $K_{11}-{K_{12}^2}/{K_{22}}$ \\
 VI & $u_{yy}=\epsilon$, $\partial \Delta F/\partial u_{xx}=0$\quad &$u_{xy}=0$ & $K_{22}-{K_{12}^2}/{K_{11}}$ \\
 VII & $u_{xy}=\epsilon/2$  \quad& $u_{xx}=u_{yy}=0$ & $K_{66}$ \\
 \hline

\end{tabular}
\caption{Types of deformations,  strain components and combined elastic moduli that contribute to the free energy change.}
\label{tab:deformation}
\end{table}
For two-dimensional structures with six-fold or four-fold symmetry,
one expects the elastic energy (\ref{eq:energy_2D}) to take
the form \cite{boal02}
\begin{equation}
\label{eq:energy_square}
\begin{aligned} 
  \Delta F/A= \frac{K_A}{2} (u_{xx} + u_{yy})^2
     + \frac{\mu_p}{2} (u_{xx} - u_{yy})^2
     + 2 \mu_s u_{xy}^2
\end{aligned},
\end{equation}
where $K_A$ is the compression modulus and $\mu_p$, $\mu_s$ are the 
pure shear and the simple shear modulus, respectively. This implies
$K_{11}=K_{22}=K_A + \mu_P$, $K_{12} = K_A - \mu_P$, and
$K_{66}=\mu_s$.  In systems with hexagonal symmetry, one additionally
expects the pure shear and the simple shear modulus to be identical,
$\mu_p = \mu_s$. Our results indeed confirm the relation
$K_{11}=K_{22}$ within the error. However, the pure shear modulus,
$\mu_p = (K_{11}-K_{12})/2 \approxeq 0.11$, differs from the simple
shear modulus, $\mu_s = K_{66} \approxeq 0.09$. We attribute
this unexpected discrepancy to nonlinear higher-order effects, 
which apparently influence the effective elastic
constants already at small deformations.  Nevertheless, since
$K_{11}=K_{22}$, the film still retains the characteristics of a
square symmetry, which explains why shear-aligning such films is so
difficult.  The Poisson ratios along the x-axis and y-axis are given
by $\nu_{x}=K_{12}/K_{22}\approxeq0.299$ and
$\nu_{y}=K_{12}/K_{11}\approxeq 0.303$, which is consistent with
Figure~\ref{fig3:stretching_el}d. 

\begin{figure}[htbp]
\begin{center}
\includegraphics[scale=1.]{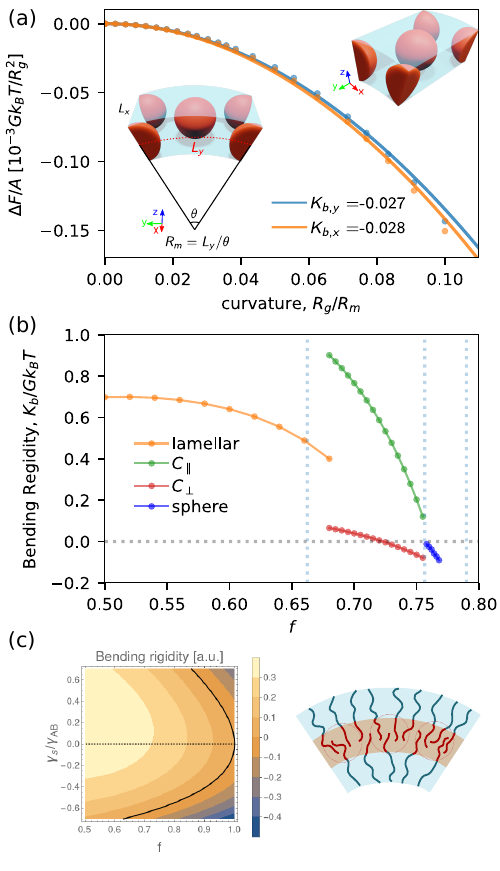}
\end{center}
\caption{(a) Free energy change per area ($F/A$) in units of $Gk_BT/R_g^2$ versus curvature ($R_g/R_m$). The solid curves are quadratic fits of the data from SCF calculations, $K_{b,x}$ and $K_{b,y}$ are fitted bending constant for bending along the  x and y-axis as show in two insets.(b) Bending rigidity as a function of volume fraction $f$ of A-block in different phases as indicated. In cylinder phases, bending deformations were applied in two directions,
with cylinders along ($C_\parallel$) and perpendicular ($C_\perp$) to the direction of curvature. The vertical dotted lines indicate bulk transition values $f=0.66, 0.76, 0.79$ for the gyroid/cylinder, cylinder/sphere and sphere/disorder transitions\cite{matsenFastAccurateSCFT2009},
(c) Bending rigidity from SST calculations for a lamellar film as a function of $f$ and ratio of surface and AB interface energy, $\gamma_s/\gamma_{AB}$; (d) Cartoon of chain rearrangement of the lamellar thin film due to bending.} 
\label{fig4:bending_elasticity}
\end{figure}

Next we study the coupling between BCP films and curvature. To this end,
we consider films that are confined between two coaxial cylinders with a
middle-surface curvature of $1/R_m$, as shown in the inset of
Fig.~\ref{fig4:bending_elasticity}. The bending free energy per unit
area of an isotropic fluid-like membrane can be described by the
Helfrich formula
\cite{helfrichElasticPropertiesLipid1973,safranCurvatureElasticityThin1999}:
$\frac{\Delta F}{A}=\frac{K_b}{2}\left(2H-c_o\right)^2+ K_g H_k$, where $K_b$ is
the bending modulus, $K_g$ is the Gaussian (or saddle-splay) modulus for
saddle-like deformations, $H=(c_1+c_2)/2$, $c_o$ and $H_k=c_1c_2$ are
the mean, spontaneous, and Gaussian curvatures, respectively, while $c_1$ and $c_2$ are the principal curvatures.
For cylindrical bending, we simply have $H=\frac{1}{2R_m}$, where
$1/R_m$ is the corresponding curvature of the middle surface of the
membranes, and the Gaussian saddle-like deformation vanishes, $H_k=0$.
Since we consider films with symmetric surface interactions, the
spontaneous curvature vanishes as well, i.e., $c_o=0$.  Therefore, the
overall bending free energy is simply a quadratic function of the
curvature, $\frac{\Delta F}{A}=\frac{1}{2}K_b \frac{1}{R_m^2}$ with the
bending constant $K_b$.

The SCFT calculations are done in cylindrical coordinates. The resulting
free energy change per surface area is shown as a function of the
curvature radius in Fig.~\ref{fig4:bending_elasticity} a.  Surprisingly,
the free energy decreases if the film is bent:  A quadratic fit to
the concave energy curves give two ``negative'' bending constants. This implies
that free-standing symmetric sphere-patterned 
monolayer films are inherently unstable, according to the SCFT predictions, different from the cylindrically patterned
monolayers which we studied previously
\cite{vuCurvatureGuidingField2018}. 
\del{
Indeed, in our experiments, we have
not been able to synthesize such free standing sphere monolayer
membranes. The experimental techniques used to make cylinder-patterned
BCP films were not successful when applied to melts of sphere-forming
BCPs.} 
The two bending constants $K_{b,11}$ and $K_{b,22}$ for bending in the $x$ and $y$ directions
are very close to each other, making it very unlikely that
curvature could be used for pattern orientation.  The BCP films show a
very slight tendency of bending in the direction of the $x$-axis,
which is also the direction of nearest neighbors. We should note that
the absolute value of $K_b$ is much smaller
for sphere-patterned BCP films than for cylinder-forming films.
Nevertheless, the change of sign leads to a fundamentally different
qualitative behavior of the films.

To investigate the origin of this behavior, we have calculated the bending rigidities 
for a large range of block fractions $f$, see Fig.\ \ref{fig4:bending_elasticity}b). 
As a function of $f$, the BCP film undergoes a sequence of structural phase 
transitions from lamellar to cylinders (C) to sphere. The bending rigidity jumps at
phase  boundaries, but apart from these discontinuities, it generally decreases with 
increasing $f$. We speculate that, as the copolymer becomes more asymmetric, the individual chains, under bending, tend to rearrange themselves to minimize the 
interfacial energies and the chain entropy. Strong stretching theory (SST) 
calculations\cite{MW_88, AL_91,Wang_92} for the lamellar structures (for derivations and equations see Supporting information S2) show that this indeed leads to a decrease of the bending constant and enables negative bending constants 
at large $f$, see Fig.~\ref{fig4:bending_elasticity}c.
\section{Conclusions}
In conclusion, we have determined the in-plane elastic moduli and the
orientation dependent bending moduli of thin symmetric
sphere-patterned films. Our results show that the film shows nearly
isotropic behavior, and in particular, the coupling between curvature
and in-plane structure is much weaker than in cylinder-forming BCP
films. This explains the experimental observation that shear-alignment
of sphere monolayer films seems almost impossible, and suggests that
curvature-based alignment strategies will also not be successful.
Most importantly, the calculations predict that the bending moduli of
sphere monolayers are negative. This implies that a free-standing single-layered BCP film in the sphere phase should inherently be unstable towards bending, potentially explaining experimental challenges (see below) in achieving either curved crystal monolayers or stable free-standing monolayers at temperatures above the glass transition temperature of both blocks.

We have explored two different sphere forming block copolymers to obtain monolayers of free-standing membranes: poly(styrene)-block-poly(ethylene-alt-propylene) and poly(styrene)-block- poly(n-hexylmethacrylate). In both cases, we found that the free standing films become unstable upon increasing the temperature well above 100$^\circ C$, transitioning the diblock copolymer into a molten state. In addition, we have also explored the stability of a monolayer of sphere forming PE-PEP block copolymer deposited via spin casting onto a curved substrate. While in this case, polymer-substrate and polymer-air interfaces has different surface energies, we found that the polymer film becomes unstable and dewets even at regions with shallow curvatures. 

However, it's important to note that Mansky et al.\cite{manskyMonolayerFilmsDiblock1995} demonstrated that free-standing sphere-forming diblock copolymer films could be achieved using polystyrene-block-polybutadiene diblocks with a spherical microstructure. This implies that film stability may not be universally predetermined and could vary based on factors such as surface tension, TEM grid size, TEM grid-polymer interactions, and the stress-strain fields introduced during sample preparation. In order to stabilize self-assembled sphere-patterned monolayer films,
special strategies will have to be applied, such as, early
crosslinking\cite{quemenerFreeStandingNanomaterialsBlock2010} , employing high molecular weight triblock copolymers or using selective and temperature-tunable solvent\cite{sohnNanopatternsFreeStandingMonolayer2001,luTemperatureTunableMicellization2010}. Alternatively, it may be possible to obtain stable films
from melts where the close-packed cubic phases have a wider range of
stability, e.g., due to additives \cite{matsen95,huang03,chen19} or
due to the effect of polydispersity of the different blocks
\cite{matsen07,matsen13,zhangEmergenceHexagonallyClosePacked2021}.

\begin{acknowledgement}
This work was funded by the Deutsche Forschungsgemeinschaft (DFG,
Germany), Grant number 248882694. We also gratefully acknowledge
financial support from the Deutsche Froschungsgemeinschaft
-- SFB 1552, Grant Number 465145163, Project C01, from the National Science Foundation MRSEC Program through the Princeton Center for 
Complex Materials (DMR-1420541), Universidad
Nacional del Sur, and the National Research Council of Argentina
(CONICET). We thank Richard Register for providing us the diblock
copolymer.
\end{acknowledgement}

\begin{suppinfo}
\section{S1: Self-consistent field Theory (SCFT)} 
Self-consistent field theory has proven to be a powerful approach for describing and predicting equilibrium structures in inhomogeneous polymer systems. In this study, we consider a melt of asymmetric AB diblock copolymer molecules with a degree of polymerization $N$, confined between two flat (coaxial cylindrical) surfaces separated by a distance of h. We assume that the majority block A takes a fraction $f$ of each diblock copolymer chain, and both blocks share the same monomer size $b$. The microscopic concentration operators of A and B segments at a given point $r$ are given by:
\begin{align}
&\hat{\phi}_A(\mathbf{r})=\frac{1}{\rho_c} \sum_{j=1}^n \int_0^f d s \delta\left(\mathbf{r}-\mathbf{r}_j(s)\right) \\
&\hat{\phi}_B(\mathbf{r})=\frac{1}{\rho_c} \sum_{j=1}^n \int_f^1 d s \delta\left(\mathbf{r}-\mathbf{r}_j(s)\right).
\end{align}
where $\rho_c=n/V$ is the average copolymer density, $n$ is the total number of copolymer molecules and $V$ is the volume of the film. Here and throughout, we will set $k_B T = 1$. The interaction potential of the melt is thus 
\begin{align}
\mathcal{H}_{\mathcal{I}}=& \rho_c \int d \mathbf{r}\left[\chi N \hat{\phi}_A(\mathbf{r}) \hat{\phi}_B(\mathbf{r})+\frac{1}{2} \kappa N \left(\hat{\phi}_A(\mathbf{r})+\hat{\phi}_B(\mathbf{r})-1\right)^2\right] \nonumber\\
&+\rho_c \int d \mathbf{r} H(\mathbf{r})\left[\Lambda_A N \hat{\phi}_A(\mathbf{r})+\Lambda_B N \hat{\phi}_B(\mathbf{r})\right]
\end{align}
where $\chi$ is the Flory-Huggins parameter specifying the repulsion of $A$ and $B$ segments. $\kappa$ is the inverse of isothermal compressibility parameter and  $\Lambda_{A, B} H(\mathbf{r})$ are surface energy fields. We assume that symmetric boundary wetting conditions, the surfaces interact energy with A and B segments are $\Lambda_A$ and $\Lambda_B$, respectively. We choose a surface field of width $\epsilon$, given by 
\begin{equation}
    H(\mathbf{r})= \begin{cases}(1+\cos (\pi z / \epsilon)) & 0 \leqslant z \leqslant \epsilon \\ 0 & \epsilon \leqslant z \leqslant h-\epsilon \\ (1+\cos (\pi(h-z) / \epsilon)) & h-\epsilon \leqslant z \leqslant h\end{cases}
\end{equation}
In our model, we choose $\epsilon=0.2 R_g$ and $R_g=b\sqrt{N}/6$ is the radius of gyration of one polymer chain. In the grand canonical ensemble, the free energy has the form
\begin{align}
{F}=&-e^\mu Q+\rho_c \int d \mathbf{r}\left[\chi N \phi_A(\mathbf{r}) \phi_B(\mathbf{r})+\frac{1}{2}\kappa N\left(\phi_A(\mathbf{r})+\phi_B(\mathbf{r})-1\right)^2\right] \nonumber\\
&-\rho_c \int d \mathbf{r}\left[\omega_A(\mathbf{r}) \phi_A(\mathbf{r})+\omega_B(\mathbf{r}) \phi_B(\mathbf{r})\right] \nonumber\\
&+\rho_c \int d \mathbf{r} H(\mathbf{r}) N\left[\Lambda_A \phi_A(\mathbf{r})+\Lambda_B \phi_B(\mathbf{r})\right]
\label{eq:free_energ}
\end{align}
where $\mu$ is chemical potential, $Q$ is the partition function of a single non-interacting polymer chain,
\begin{equation}
    Q=\int d \mathbf{r} q(\mathbf{r}, s) q^{\dagger}(\mathbf{r}, 1-s)
\end{equation}
and end-segment distribution functions $q(\mathbf{r}, s)$ and  $q^{\dagger}(\mathbf{r}, 1-s)$ satisfy the modified diffusion equation
\begin{equation}
    \frac{\partial q(\mathbf{r}, s)}{\partial s}=\Delta q(\mathbf{r}, s)-\omega_\alpha(\mathbf{r}, s) q(\mathbf{r}, s)
\end{equation}
with
\begin{equation}
    \omega_\alpha(\mathbf{r}, s)= \begin{cases}\omega_A(\mathbf{r}) & \text { for } 0<s<f \\ \omega_B(\mathbf{r}) & \text { for } f<s<1\end{cases}
\end{equation}
and the initial condition $q(\mathbf{r}, 0)=1$. The diffusion equation for $q^{\dagger}(\mathbf{r}, 1-s)$ is similar with $\omega_\alpha(\mathbf{r}, s)$ replaced by $\omega_\alpha(\mathbf{r}, 1-s)$ and the same initial condition, $q^{\dagger}(\mathbf{r}, 0)=1$. By finding the extremum of the free energy with respect to $\omega_{A, B}(\mathbf{r})$ and $\phi_{A, B}(\mathbf{r})$, we get the self-consistent equations,



\begin{align}
\frac{\omega_A(\mathbf{r})}{N} &=\chi \phi_B(\mathbf{r})+\kappa\left[\phi_A(\mathbf{r})+\phi_B(\mathbf{r})-1\right]+\Lambda_A H(\mathbf{r}) \\
\frac{\omega_B(\mathbf{r})}{N} &=\chi \phi_A(\mathbf{r})+\kappa\left[\phi_A(\mathbf{r})+\phi_B(\mathbf{r})-1\right]+\Lambda_B H(\mathbf{r}) \\
\phi_A(\mathbf{r}) &=\frac{1}{\rho_c} \mathrm{e}^\mu \int_0^f d s q(\mathbf{r}, s) q^{\dagger}(\mathbf{r}, 1-s) \\
\phi_B(\mathbf{r}) &=\frac{1}{\rho_c} \mathrm{e}^\mu \int_f^1 d s q(\mathbf{r}, s) q^{\dagger}(\mathbf{r}, 1-s)
\end{align}

\section{S2: Strong-stretching theory (SST) }
\begin{figure}[htbp]
\begin{center}
\includegraphics[scale=0.5]{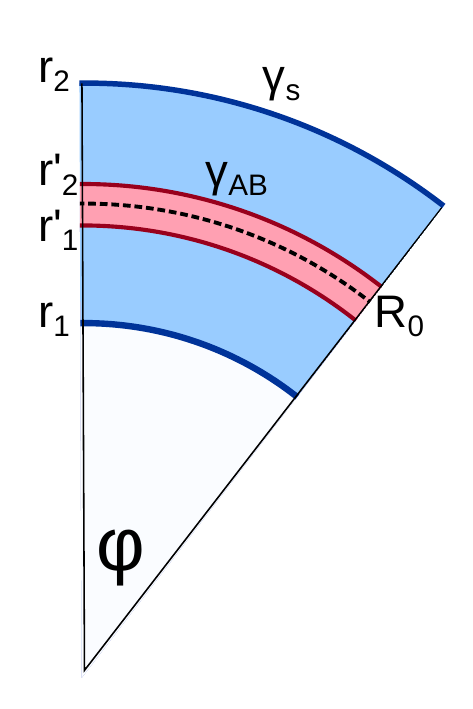}
\end{center}
\caption{Cartoon of the geometrical setting considered in the SSL calculations (cylindrical case). We consider two incompressible block copolymer layers facing each other with their B blocks, with the contact surface located at the radius $R_0$ and the outer surfaces radii at $r_1$ and $r_2$. The interfaces between A blocks (blue) and B blocks (pink) are located at radii $r_1'$ and $r_2'$.} 
\label{fig:ssl_cartoon}
\end{figure}

In the SST calculations, we consider the cylindrical geometry sketched in Fig. \ref{fig:ssl_cartoon}, and, for completeness, the corresponding spherical geometry. Two copolymer layers (labeled $i = 1,2$) face each other at the radius $R_0$. The two outer surfaces of the film are found at the radii $r_1$ and $r_2$. The thickness of the film is thus $D=(r_2-r_1)$, and the bilayer midplane is located at $R_{\textrm{mid}}=\frac{1}{2}(r_1+r_2)$. Our goal is to minimize the grand canonical free energy per area as a function of $R_{\textrm{mid}}$ with respect to $R_0$ and $D$ and expand it in powers of $1/R_{\textrm{mid}}$. This will allow us to calculate the bending rigidity $K_b$ and the Gaussian rigidity $K_G$ in the free energy expression.
\begin{equation}
\label{aeq:Helfrich}
F = F_{\textrm{planar}} + \int \textrm{d}A \: \Big(\frac{1}{2} K_b  \: (2H)^2
+ K_G \: K \Big)
\end{equation}
for a general curved surface with total curvature $2H=c_1+c_2$ and Gaussian curvature $K = c_1 c_2$ (the $c_i$ are the principal curvatures).

To proceed, we first calculate the canonical (Helmholtz) free energy. 
In SST approximation, it is given as the sum $F_{\textrm{H}}= \sum_{i} (F_{\textrm{interface}}{(i)}+F_{\textrm{stretch}}{(i)})$ of the interfacial energies and stretching energies in the system\cite{matsen01}. 
The interfacial energy per area $A_0$ at the radius $R_0$ is given 
\begin{eqnarray*}
\label{aeq:int}
F_{\textrm{interface, cyl}}^{(i)}/A_0 &=&  \gamma_s \: r_i/R_0 + \gamma_{AB} r_i'/R_0 \\
F_{\textrm{interface, sph}}^{(i)}/A_0 &=&  \gamma_s \: (r_i/R_0)^2 + \gamma_{AB} (r_i'/R_0)^2
\end{eqnarray*}
where $\gamma_s$ is the surface free energy per area of the film and   $\gamma_{AB}$ the interfacial free energy between
A and B domains ($\gamma_{AB} = G \sqrt{\chi N}/R_g^2$ in the Strong-stretching limit). 

The total stretching energy in SST approximation can be calculated from \cite{matsen01}
\begin{eqnarray*}
\label{aeq:stretch}
F_{\textrm{stretch, cyl}}^{(i)}/A_0 &=& \frac{\alpha}{f^2} \int_{r_i}^{r'_i} \textrm{d}r \:\frac{r}{R_0} \:(r-r'_i)^2
  + \frac{\alpha }{(1-f)^2} \int_{r'_i}^{R_0} \textrm{d}r\: \frac{r}{R_0} \:(r-r'_i)^2
  \\
 F_{\textrm{stretch, sph}}^{(i)}/A_0 &=& \frac{\alpha}{f^2} \int_{r_i}^{r'_i} \textrm{d}r \:(\frac{r}{R_0})^2 \:(r-r'_i)^2
  + \frac{\alpha}{(1-f)^2} \int_{r'_i}^{R_0} \textrm{d}r\: (\frac{r}{R_0})^2 \:(r-r'_i)^2 
\end{eqnarray*}
with $\alpha = \frac{\pi^2}{16} G/R_g^5$. Since the copolymers are incompressible, the ratio of volumes $V_A$ and $V$ occupied by A monomers and all monomers is $V_A/V= f$. This results in the cylindrical case in the constraints $|r_i^2 - {r'_i}^2| = f | R_0^2 - r_i^2|$ for $i=1,2$, and in the spherical case
$|r_i^3 - {r'_i}^3| = f | R_0^3 - r_i^3|$. To account for them, we introduce the variable $d_{i,\textrm{cyl}} = |R_0^2 - r_i^2|/(2 R_0)$ and $d_{i,\textrm{sph}} = |R_0^3 - r_i^3|/(2 R_0^2)$, which is roughly (not exactly) the  thickness of the monolayer $i$. In terms of $d_i$, taking into account the constraints, the radii $r_i$, $r'_i$ can be written as
\begin{displaymath}
    r_{1,2} = R_0 \: (1 \mp \nu d_{1,2}/R_0)^ {1/\nu}, \quad 
    r'_{1,2} = R_0 \: (1 \mp \nu (1-f) d_{1,2}/R_0)^{1/\nu}
\end{displaymath}
with $\nu=2$ in the cylindrical case and $\nu=3$ in the spherical case.
We insert these expressions in the above equations for the free energy and expand the result in powers of $1/R_0$, which is a small parameter for weakly bent membranes. The result has the form
\begin{displaymath}
\begin{array}{lclclcl}
F_{\textrm{stretch}}^{(1,2)}/A_0 & = &d_i^3 \: a  & \pm& \frac{1}{R_0} \: d_i^4 \: b(f) &+& \frac{1}{R_0^2} \: d_i^5 \: c(f) \\
F_{\textrm{interface}}^{(1,2)}/A_0 &=& \bar{\gamma} &\mp& \frac{1}{R_0} \: d_i \: \bar{\gamma}_1 &-& \frac{1}{R_0^2} \: d_i^2 \: \bar{\gamma}_2 \end{array}
 \end{displaymath}
 with the coefficients
 \begin{equation}
 \begin{array}{ll}
 a  = \frac{\alpha}{3}, \qquad &\bar{\gamma} = \gamma_s + \gamma_{AB} 
 \\
 b_{\textrm{cyl}}(f) = \frac{\alpha}{12} \: (5 - 2 f), \qquad & b_{\textrm{sph}}(f) = \frac{\alpha}{6} \: (5 - 2 f)
 \\
 c_{\textrm{cyl}}(f) = \frac{\alpha}{12} \: (7 - 5 f + f^2), \qquad &c_{\textrm{sph}}(f) = \frac{\alpha}{30} \: (61 - 43 f + 8f^2)
  \\
\bar{\gamma}_{1,\textrm{cyl}}(f) = \gamma_s + (1-f) \gamma_{AB}, \quad &\bar{\gamma}_{1,\textrm{sph}}(f) = 2(\gamma_s + (1-f) \gamma_{AB})
\\
\bar{\gamma}_{2,\textrm{cyl}}(f) = \frac{1}{2}(\gamma_s + (1-f)^2\gamma_{AB}), \quad & \bar{\gamma}_{2,\textrm{sph}}(f) = \gamma_s + (1-f)^2\gamma_{AB}.
\end{array}
\end{equation}
The contact radius $R_0$ may deviate from the mid radius of the bilayer, $R_{\textrm{mid}}=\frac{1}{2}(r_1+r_2)$,   the reference plane for the definition of the bending rigidity. Therefore, in the next step,  we rewrite the total free energy per area $A$ at the mid radius $ R_{\textrm{mid}}$ as an expansion in powers of $1/R_{\textrm{mid}}$, for given   total thickness $D=r_2-r_1$, and minimize it with respect to $R_0$. To this end, we define a set of reduced quantities, 
$\varepsilon=D/(2 R_{\textrm{mid}})$ (the new small
parameter), $\delta   = (R_{\textrm{mid}}/R_0 - 1)/\varepsilon$ (accounting for the fact that $R_{\textrm{mid}}/R_0 = 1$ as $\varepsilon \to 0$), and $\eta(D) = \alpha (\frac{D}{2})^3/\gamma_{AB}$. Using these quantities, minimizing $F_{\textrm{H}}/A$ with respect to $\delta$ at fixed $D$, and keeping only
terms up to order $\varepsilon$, we get
\begin{equation}  
    \frac{F_{\textrm{H}}}{A} = 2  \:\big( \gamma_s + \gamma_{AB} + \frac{1}{3} \gamma_{AB} \:  \eta(D)   \big)  
    + \kappa(f,D) \: \varepsilon^2
\end{equation}
with
\begin{eqnarray}
    \kappa_{\textrm{cyl}}(f,D) &=&  \gamma_{AB} \: f \: \Big[ 1-\frac{f}{3} 
    - \frac{f}{2 \eta(D)}  
    - \frac{1}{6}(1 + \frac{f}{3}) \:  \eta(D)
    \Big],
    \\
        \kappa_{\textrm{sph}}(f,D) &=&  \: \gamma_{AB} \: \Big[   
    2(1+ \frac{\gamma_s}{\gamma_{AB}}) + \frac{2 f^ 2}{3}  - \frac{2 f^2}{\eta(D)} 
    + \frac{\eta(D)}{45}(3 - 9 f -16 f^2)  
    \Big].
\end{eqnarray}
in the cylindrical and spherical geometry, respectively. From these two results, we can calculate the bending rigidity $K_b^{(0)}$ and the Gaussian rigidity $K_g(D)^ {(0)}$ for films of fixed thickness $D$ via
$K_b^{(0)} = \frac{1}{2} \: (\frac{D}{2})^2 \: \kappa_{\textrm{cyl}}(D)$ and
$K_G^{(0)} =  (\frac{D}{2})^2 (\kappa_{\textrm{sph}}(D) - 4 \kappa_{\textrm{cyl}}(D))$. The result is
\begin{eqnarray}
K_b^{(0)}(f,D) &=& \frac{D^2}{2} \: \gamma_{AB} \: f \:
\Big( 1-\frac{f}{3} - \frac{f}{2 \eta(D)} + \frac{f+3}{18} \: \eta(D) \Big)
\\
K_G^{(0)}(f,D) &=& \frac{D^2}{2} \: \gamma_{AB} \: 
\Big( (1-f)^ 2 + \frac{\gamma_s}{\gamma_{AB}} + \frac{\eta(D)}{30} 
(1+7 f - 2 f^2) \Big)
\end{eqnarray}

Finally, we switch to the grand canonical ensemble and minimize the grand canonical free energy
\begin{equation} 
    \frac{F(f,D)}{A} = \frac{F_{\textrm{H}}(f,D)}{A}- \mu \: \frac{\rho_0}{N} \: \frac{V(D)}{A}.
\end{equation}
with respect to $D$. Here $\mu$ is the chemical potential, $\rho_0$ is the monomer density, and $V(D)/A$ is the film volume per area $A$. We consider a general curved geometry with total curvature $2H$ and mean curvature $K$, assuming that both are small, i.e., $(2 H)^2 \propto \xi^2$ and $K \propto \xi^2$ with $\xi \ll 1$.
The general expression for the Helmholtz free energy per area of a film with given thickness $D$ is given by
\begin{equation}  
    \frac{F_{\textrm{H}}(f,D)}{A} = 2  \:\big( \gamma_s + \gamma_{AB} + \frac{1}{3} \gamma_{AB} \:  \eta(D)   \big)  
    + \frac{1}{2} \: K_b^ {(0)}(f,D) \: (2H)^2 + K_G^ {(0)}(f,D) \: K. 
\end{equation}
The volume per area can be calculated from
\begin{equation}
    \frac{V(D)}{A} =  \int_{-D/2}^{D/2} \textrm{d}z \: (1 + 2 H \: z + K \: z^2)
    =  D  \: (1+K \: D^2/12),
\end{equation}
We minimize $F(f,D)/{A}$ with respect to $D$ and expand again in powers of the small parameter, $\xi$. In the planar case, the minimization gives
\begin{equation}
    D_{\textrm{p}}(\mu) = 2 \sqrt{\frac{\rho_0}{N} \: \mu \: \frac{1}{\alpha}}
      = \frac{8}{\pi} \: R_g \sqrt{\mu}
\end{equation}
and the corresponding planar free energy is
\begin{equation}
\label{eq:f_planar}
       \frac{F_{\textrm{p}}(\mu)}{A} =     
    2 (\gamma_s + \gamma_{AB}) - \frac{16}{3 \pi} \: \frac{G}{R_g^2} \: \mu^{3/2}.
\end{equation}

In curved films, the optimal thickness deviates from $D_0$ by an amount which is proportional to $\xi^2$, $D - D_0 = D_2 \xi^2 + {\cal O}(\xi^4)$. Since $F(f,D_0)$ is a minimum with respect to $D$, this correction only contributes to quadratic order, hence order $\xi^4$, to $F(f,D)$, and can be neglected. Thus we obtain 
 \begin{equation}  
    \frac{F(\mu)}{A} = \frac{F_{\textrm{p}}(\mu)}{A}
    + \frac{1}{2} \: K_b^ {(0)}(f,D_{\textrm{p}}(\mu)) \: (2H)^2 
    + \big(K_G^{(0)}(f,D_{\textrm{p}}(\mu))-\mu \frac{\rho_0}{N} \: \frac{D_{\textrm{p}}(\mu)^3}{12} \big) \: K. 
\end{equation}
Following Ajdari and Leibler\cite{AL_91}, we now consider specifically the chemical potential 
$\mu_0$ where the planar film just becomes stable, i.e., $F_{\textrm{planar}}(\mu_0)=0$, which implies
\begin{equation}
    \mu_0 = \Big[\frac{3 \pi}{8} \: \frac{R_g^2}{G} \: (\gamma_s + \gamma_{AB}) \Big]^{2/3} = \frac{3 \: \pi^{2/3}}{2^{5/3}}  \:
    \big( (\gamma_s + \gamma_{AB})^2 \: N \big)^{1/3} \: (\rho_0 b)^{-2/3}.
\end{equation}
(with the statistical segment length $b$). The latter expression is also given in Ref.\ \cite{AL_91} for $\gamma_s = 0$. At this value of $\mu_0$, the thickness $D_0=D_{\textrm{planar}}(\mu_0)$ and in particular the reduced parameter $\eta_0 = \eta(D_0)$ take the simple form
\begin{equation}
\label{aeq:D_opt}
\eta_0 = \frac{3}{2}(1+ \frac{\gamma_s}{\gamma_{AB}}),  \qquad
    D_0 = 2 (\frac{3}{2 \alpha})^{1/3} \: (\gamma_s + \gamma_{AB})^{1/3}
\end{equation}
Inserting this in the equations for $K_b^{(0)}$ and $K_G^{(0)}$, we finally 
obtain the following expressions for the bending rigidity and the effective Gaussian rigidity:
\begin{eqnarray}
\label{aeq:kb}
K_b &=& \frac{D_0^2}{2} \: \gamma_{AB} \: f \: \Big( \frac{3}{4}(1-f) - \frac{f+3}{12} \: \frac{\gamma_s}{\gamma_{AB}} + \frac{f}{3} \: \frac{\gamma_s}{\gamma_s + \gamma_{AB}} \Big)
\\
\label{aeq:kg}
K_G &=& \frac{D_0^2}{2} \: \gamma_{AB} \: \Big( f(f-2) + \frac{1}{20} \:  (1+\frac{\gamma_s}{\gamma_{AB}}) (11 + 7 f - 2 f^2) \Big),
\end{eqnarray}
in the free energy expansion (\ref{aeq:Helfrich}) (at $F_{\textrm{planar}}=0$).
In the special case $\gamma_s=0$, these equations reproduce the result of Ajdari and Leibler\cite{AL_91} (taking into account that these authors define $f$ as $1-f$ in our notation). Inspecting the free energy expression (\ref{aeq:Helfrich}) reveals two stability conditions for a flat bilayer: $K_G < 0$ and $2 K_b + K_G>0$. Figure \ref{fig:ssl1} shows the rescaled results for $K_b$ and $K_c$ as a function of $f$ and $\gamma_s/\gamma_{AB}$, also indicating this stability region. The rescaling constant is $(D_0/2)^2 \: \gamma_{AB}$.

\begin{figure}[htbp]
\begin{center}
\includegraphics[scale=0.45]{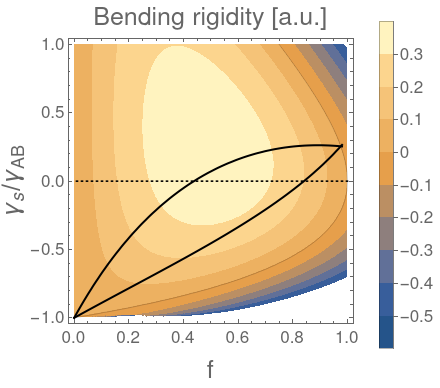}
\includegraphics[scale=0.45]{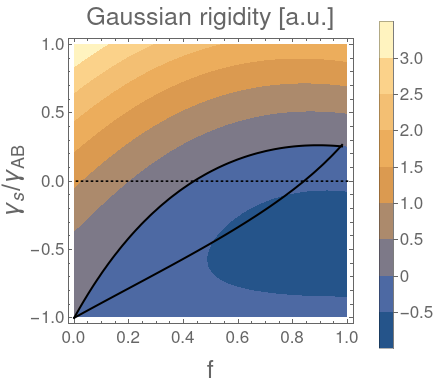}
\end{center}
\caption{Rescaled bending rigidity $K_b$ (left) and Gaussian rigidity $K_G$ for lamellar films in the strong stretching limit versus $A$ block copolymer ratio $f$ and ratio of interfacial energies $\gamma_s/\gamma_{AB}$. The thick solid lines surround the parameter region where the bilayer is stable with respect to bending deformations. 
\label{fig:ssl1}}
\end{figure}

In the lamellar state, for $f > 0.5$, the bending rigidity generally decreases with $f$, and eventually becomes negative for all values of surface tension except $\gamma_s \equiv 0$. To rationalize this behavior, it is instructive to investigate the  asymmetry of the bent film in the cylindrical geometry, i.e., the extent to which the contact surface between the monolayers, located at $R_0$, deviates from the mid surface of the bilayer, $R_{\textrm{mid}}$. We define
the asymmetry parameter $\Delta$ via $R_0 = R_{\textrm{mid}} \: (1+\Delta \: \varepsilon^2)$ (recalling $\varepsilon - D/2 R_{\textrm{mid}}$). Its value is obtained as
\begin{equation}
    \Delta(f) = \frac{1}{3} \: f \frac{\gamma_s}{\gamma_s + \gamma_{AB}},
\end{equation}
and shown in Figure \ref{fig:ssl}(right). The asymmetry of the film results from two opposing driving 
factors: At fixed $D$, the interfacial energy due to AB incompatibility ($\gamma_{AB}$) can be reduced 
if $R_0$ is driven inwards, towards smaller values, because this reduces the total AB interfacial
area. (At constant $R_{\textrm{mid}}$, the total outer surface area is constant in the cylindrical geometry).
On the other hand, the stretching energy benefits if $R_0$ is driven outwards: Single copolymers in the
inner monolayer are more stretched (lose energy), whereas single copolymers in the outer monolayer are
less stretched (gain energy). Since the outer layer contains more copolymers, the net effect is positive
for $R_0> R_{\textrm{mid}}$. This latter factors dominates if the film thickness $D$ is large, i.e., $\gamma_s+\gamma_{AB}$ is large. Then we find $R_0 > R_{\textrm{mid}}$. If $D$ is small, i.e., $\gamma_s+\gamma_{AB}$ is small, the  effect of the interfacial energy dominates, and we find $R_0 < R_{\textrm{mid}}$. It is interesting to note that the two effects exactly compensate each other in the case $\gamma_s=0$. In that case, $R_0 = R_{\textrm{mid}}$
and the bending rigidity is always positive.  
 
\begin{figure}[htbp]
\begin{center}
\includegraphics[scale=0.45]{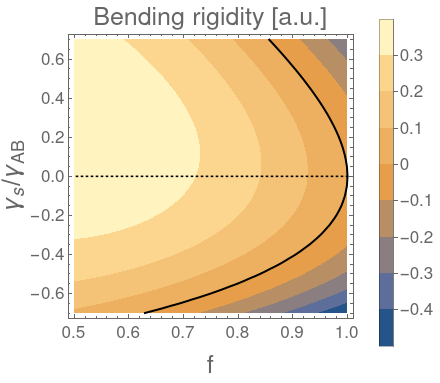}
\includegraphics[scale=0.45]{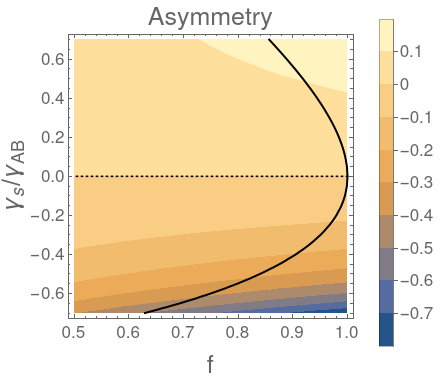}
\end{center}
\caption{Left: Rescaled Bending rigidity for lamellar films in the strong stretching limit versus $A$ block copolymer ratio $f$ and ratio of interfacial energies $\gamma_s/\gamma_{AB}$ in the regime $f > 0.5$. 
Right: Corresponding asymmetry parameter $\Delta$ (see text for explanation). The solid lines shows the parameter combinations for which $K_b \equiv 0$.
\label{fig:ssl}}
\end{figure}

Finally, we note that instead of minimizing the grand canonical free energy per area with respect to the film thickness $D$, we could also minimize the canonical Helmholtz free energy per chain, i.e., per volume. In the planar case, this results in the same optimal thickness $D_0$ (Eq.\ (\ref{aeq:D_opt})) than before, and the Helmholtz free per chain reads 
\begin{equation}
    \frac{F_{\textrm{H}}}{V} = \frac{1}{D_0} \Big( 3 (\gamma_s + \gamma_{AB}) + \frac{1}{2} K_B \: (2H)^2 + K_G \: K\Big), 
\end{equation}
where $K_B$ and $K_G$ are given by Eqs.\ (\ref{aeq:kb}), (\ref{aeq:kg}). Hence the results from the two approaches are consistent. 

\section{S3: Numerical Details of ADI Method}
The SCFT calculations are done in Cartesian and Cylindrical coordinates with Dirichlet boundary conditions in the direction normal to the surface and periodic boundary conditions in two in-plane directions. The modified diffusion equation is solved numerically using the Crank-Nicolson scheme and alternative direction implicit (ADI) method.
\subsection{3D ADI Algorithm:~Cartesian Coordinate}

The three dimensional diffusion equation of the propagator in Cartesian coordinate has the form 

\begin{equation}
    \frac{\partial q(\mathbf{r}, s)}{\partial s}=\left(\frac{\partial^2}{\partial x^2}+\frac{\partial^2}{\partial y^2}+\frac{\partial^2}{\partial z^2}\right) q(\mathbf{r}, s)-\omega(\mathbf{r}, s) q(\mathbf{r}, s)\\
    \label{apeq:3D_diffusion}
\end{equation}

In order to solve eq.~\ref{apeq:3D_diffusion}, we apply the Crank-Nicolson scheme in using Douglas-Gunn ADI scheme by introducing two intermediate variables ${q}^{n+1/3}$ and ${q}^{n+2/3}$. The discretization of eq.~\ref{apeq:3D_diffusion} is done by splitting into three subset of equations:

\begin{align}
\frac{{q}_{i, j, k}^{n+1/3}-q_{i, j, k}^n}{\Delta s}=& \frac{1}{2}\left(\frac{\delta_x^2}{\Delta x^2}-\frac{\omega_{i, j, k}}{3}\right)\left(q_{i, j, k}^{n+1/3}+q_{i, j, k}^n\right) \nonumber\\
&+\left(\frac{\delta_y^2}{\Delta y^2}+\frac{\delta_z^2}{\Delta z^2}-2 \frac{\omega_{i, j, k}}{3}\right) q_{i, j, k}^n \\
\frac{q_{i, j, k}^{n+2/3}-q_{i, j, k}^n}{\Delta s}=& \frac{1}{2}\left(\frac{\delta_x^2}{\Delta x^2}-\frac{\omega_{i, j, k}}{3}\right)\left(q_{i, j, k}^{n+1/3}+q_{i, j, k}^n\right) \nonumber\\
&+\frac{1}{2}\left(\frac{\delta_y^2}{\Delta y^2}-\frac{\omega_{i, j, k}}{3}\right)\left(q_{i, j, k}^{n+2/3}+q_{i, j, k}^n\right) \nonumber\\
&+\left(\frac{\delta_z^2}{\Delta z^2}-\frac{\omega_{i, j, k}}{3}\right) q_{i, j, k}^n \\
\frac{q_{i, j, k}^{n+1}-q_{i, j, k}^n}{\Delta s}=& \frac{1}{2}\left(\frac{\delta_x^2}{\Delta x^2}-\frac{\omega_{i, j, k}}{3}\right)\left(q_{i, j, k}^{n+1/3}+q_{i, j, k}^n\right) \nonumber\\
&+\frac{1}{2}\left(\frac{\delta_y^2}{\Delta y^2}-\frac{\omega_{i, j, k}}{3}\right)\left(q_{i, j, k}^{n+2/3}+q_{i, j, k}^n\right) \nonumber\\
&+\frac{1}{2}\left(\frac{\delta_z^2}{\Delta z^2}-\frac{\omega_{i, j, k}}{3}\right)\left(q_{i, j, k}^{n+1}+q_{i, j, k}^n\right)
\end{align}
where $i=\left\{1, N_x\right\}$ denotes the grid points in $x$-direction, $j=\left\{1, N_y\right\}$ denotes the grid points in the  $y$-direction, and $k=\left\{1, N_z\right\}$ denotes the grid points in the $z$-direction. $\Delta x$, $\Delta y$, and $\Delta z$ are interval sizes in space. Using the second-order central difference algorithm, the second derivative with respect to $x$ can be approximated as 

\begin{equation}
    \frac{\partial^2 f_{i, j, k}}{\partial x^2} \approx \frac{\delta_x^2 f_{i, j,k}}{\Delta x^2}=\frac{f_{i+1, j,k}-2 f_{i, j,k}+f_{i-1, j,k}}{\Delta x^2}.
\end{equation}

Approximate and rearrange above equations, we obtain the three-step ADI 
\begin{align}
\frac{-\Delta s}{2 \Delta x^2} q_{i-1, j, k}^{n+1/3}+\left(1+\frac{\Delta s}{\Delta x^2}+\frac{\Delta s \omega_{i, j, k}}{6}\right) q_{i, j, k}^{n+1/3}-\frac{\Delta s}{2 \Delta x^2} q_{i+1, j, k}^{n+1/3} \nonumber\\
=\frac{\Delta s}{2 \Delta x^2} q_{i-1, j, k}^n+\frac{\Delta s}{\Delta y^2} q_{i, j-1, k}^{n}+\frac{\Delta s}{\Delta z^2} q_{i, j, k-1}^{n} \nonumber\\
+\frac{\Delta s}{2 \Delta x^2} q_{i+1, j, k}^n+\frac{\Delta s}{\Delta y^2} q_{i, j+1, k}^n+\frac{\Delta s}{\Delta z^2} q_{i, j, k+1}^n \nonumber\\
+\left(1-\frac{\Delta s}{\Delta x^2}-\frac{2 \Delta s}{\Delta y^2}-\frac{2 \Delta s}{\Delta z^2}-\frac{5 \Delta s \omega_{i, j, k}}{6}\right) q_{i, j, k}^n
\end{align}

\begin{align}
\frac{-\Delta s}{2 \Delta y^2} q_{i, j-1, k}^{n+2/3}+\left(1+\frac{\Delta s}{\Delta y^2}+\frac{\Delta s \omega_{i, j, k}}{6}\right) q_{i, j, k}^{n+2/3}-\frac{\Delta s}{2 \Delta y^2} q_{i, j+1, k}^{n+2/3} 
\nonumber\\=\frac{-\Delta s}{2 \Delta y^2} q_{i, j-1, k}^n+q_{i, j, k}^{n+1/3}+\left(\frac{\Delta s}{\Delta y^2}+\frac{\Delta s \omega_{i, j, k}}{6}\right) q_{i, j, k}^n-\frac{\Delta s}{2 \Delta y^2} q_{i, j+1, k}^n
\end{align}

\begin{align}
&\frac{-\Delta s}{2 \Delta z^2} q_{i, j, k-1}^{n+1}+\left(1+\frac{\Delta s}{\Delta z^2}+\frac{\Delta s \omega_{i, j, k}}{6}\right) q_{i, j, k}^{n+1}-\frac{\Delta s}{2 \Delta z^2} q_{i, j, k+1}^{n+1} \nonumber\\
&=\frac{-\Delta s}{2 \Delta z^2} q_{i, j, k-1}^n+q_{i, j, k}^{n+2/3}+\left(\frac{\Delta s}{\Delta z^2}+\frac{\Delta s \omega_{i, j, k}}{6}\right) q_{i, j, k}^n-\frac{\Delta s}{2 \Delta z^2} q_{i, j, k+1}^n
\end{align}
\subsection{3D ADI:~Cylindrical coordinates}
The modified diffusion equation in three Dimensional Cylindrical Coordinates reads;
\begin{equation}
\label{eq:diff_cylin}
    \frac{\partial q(r, \theta,z,s)}{\partial s}=\left(\frac{\partial_r^2}{\partial r^2}+\frac{1}{r}\frac{\partial_r}{\partial r}+\frac{1}{r^2}\frac{\partial_\theta^2}{\partial\theta^2}+\frac{\partial_z^2}{\partial_z z^2}\right)q(r,\theta,z,s)-\omega(r,\theta,z,s)q(r,\theta,z,s)
\end{equation}
Applying the ADI method to the CN scheme, we split the calculation into 3 sub-steps to implicitly solve eq.\,\ref{eq:diff_cylin} in $r$, $\theta$ and $z$ directions at step $n+1/3$, $n+2/3$ and $n+1$, The discretization of eq.\, \ref{eq:diff_cylin} gives 
\begin{align}
\label{eq:cylin_1}
    \frac{q_{i,j,k}^{n+1/3}-q_{i,j,k}^{n}}{\Delta s}=&\left(\frac{\delta_r^2}{\Delta r^2}+\frac{1}{r_i}\frac{\delta_r}{\Delta r}-\frac{\omega_{i,j,k}}{3}\right)\frac{(q_{i,j,k}^{n+1/3}+q_{i,j,k}^n)}{2}\\
    &\nonumber+\left(\frac{1}{r_i^2}\frac{\delta_\theta^2}{\Delta \theta^2}+\frac{\delta_z^2}{\Delta z^2}-\frac{2}{3}\omega_{i,j,k}\right)q_{i,j,k}^n,
\end{align}
\begin{align}
\label{eq:cylin_2}
    \frac{q_{i,j,k}^{n+2/3}-q_{i,j,k}^{n}}{\Delta s}=&\left(\frac{\delta_r^2}{\Delta r^2}+\frac{1}{r_i}\frac{\delta_r}{\Delta r}-\frac{\omega_{i,j,k}}{3}\right)\frac{(q_{i,j,k}^{n+1/3}+q_{i,j,k}^n)}{2}\\
    &\nonumber +\left(\frac{1}{r_i^2}\frac{\delta_\theta^2}{\Delta \theta^2}-\frac{\omega_{i,j,k}}{3}\right)\frac{(q_{i,j,k}^{n+2/3}+q_{i,j,k}^n)}{2}\\
    &\nonumber+\left(\frac{\delta_z^2}{\Delta z^2}-\frac{\omega_{i,j,k}}{3}\right)q_{i,j,k}^n
\end{align}
\begin{align}
\label{eq:cylin_3}
    \frac{q_{i,j,k}^{n+1}-q_{i,j,k}^{n}}{\Delta s}=&\left(\frac{\delta_r^2}{\Delta r^2}+\frac{1}{r_i}\frac{\delta_r}{\Delta r}-\frac{\omega_{i,j,k}}{3}\right)\frac{(q_{i,j,k}^{n+1/3}+q_{i,j,k}^n)}{2}\\
    &\nonumber +\left(\frac{1}{r_i^2}\frac{\delta_\theta^2}{\Delta \theta^2}-\frac{\omega_{i,j,k}}{3}\right)\frac{(q_{i,j,k}^{n+2/3}+q_{i,j,k}^n)}{2}\\
    &\nonumber+\left(\frac{\delta_z^2}{\Delta z^2}-\frac{\omega_{i,j,k}}{3}\right)\frac{(q_{i,j,k}^{n+1}+q_{i,j,k}^n)}{2}
\end{align}
where $i\in [1,N_r]$, $j\in[1,N_\theta]$, and $k\in[1,N_z]$, are the grid points in $r$, $\theta$ and $z$ directions, respectively. Again using the second-order central difference algorithm, the $\delta$ operator is simply,
\begin{align}
    &\delta^2_\gamma f=f_{\gamma-1}-2f_\gamma+f_{\gamma+1}\\
    &\delta_\gamma f=f_{\gamma+1}-f_{\gamma-1}
\end{align}
Combining and rearranging eqs. \ref{eq:cylin_1}-\ref{eq:cylin_3}, we have
\begin{align}
    &\left(-\frac{\Delta s}{2\Delta r^2} +\frac{\Delta s}{2r_i\Delta r}\right)q_{i-1,j,k}^{n+1/3}+
    \left(1+\frac{\Delta s}{\Delta r^2}+\frac{\Delta s}{6}\omega_{i,j,k}\right)q_{i,j,k}^{n+1/3}-
    \left(\frac{\Delta s}{2\Delta r^2} +\frac{\Delta s}{2r_i\Delta r}\right)q_{i+1,j,k}^{n+1/3}\\
    &=\nonumber\left(\frac{\Delta s}{2\Delta r^2} -\frac{\Delta s}{2r_i\Delta r}\right)q_{i-1,j,k}^{n}+\left(1-\frac{\Delta s}{\Delta r^2}-\frac{5\Delta s}{6}\omega_{i,j,k}-\frac{2\Delta s}{r_i^2\Delta\theta^2}-\frac{2\Delta s}{\Delta z^2}\right)q_{i,j,k}^{n}\\
    &\nonumber+\left(\frac{\Delta s}{2\Delta r^2} +\frac{\Delta s}{2r_i\Delta r}\right)q_{i+1,j,k}^{n}+\frac{\Delta s}{r_i^2\Delta \theta^2}\left(q_{i,j-1,k}^n+q_{i,j+1,k}^n\right)+\frac{\Delta s}{\Delta z^2}\left(q_{i,j,k-1}^n+q_{i,j,k+1}^n\right)
\end{align}
\begin{align}
    &-\frac{\Delta s}{2r_i^2\Delta \theta^2}q_{i,j-1,k}^{n+2/3}+
    \left(1+\frac{\Delta s}{r_i^2\Delta \theta^2}+\frac{\Delta s}{6}\omega_{i,j,k}\right)q_{i,j,k}^{n+2/3}-\frac{\Delta s}{2r_i^2\Delta \theta^2}q_{i,j+1,k}^{n+2/3}\\
    &=\nonumber q_{i,jk}^{n+1/3}-\frac{\Delta s}{2r_i^2\Delta \theta^2}q_{i,j-1,k}^{n}+\left(\frac{\Delta s}{r_i^2\Delta \theta^2}+\frac{\Delta s}{6}\omega_{i,j,k}\right)q_{i,j,k}^{n}-\frac{\Delta s}{2r_i^2\Delta \theta^2}q_{i,j+1,k}^{n}
\end{align}
\begin{align}
    &-\frac{\Delta s}{2\Delta z^2}q_{i,j,k-1}^{n+1}+
    \left(1+\frac{\Delta s}{\Delta z^2}+\frac{\Delta s}{6}\omega_{i,j,k}\right)q_{i,j,k}^{n+1}-\frac{\Delta s}{2\Delta z^2}q_{i,j,k+1}^{n+1}\\
    &=\nonumber q_{i,jk}^{n+2/3}-\frac{\Delta s}{2\Delta z^2}q_{i,j,k-1}^{n}+\left(\frac{\Delta s}{\Delta z^2}+\frac{\Delta s}{6}\omega_{i,j,k}\right)q_{i,j,k}^{n}-\frac{\Delta s}{2\Delta z^2}q_{i,j,k+1}^{n}
\end{align}

\subsection{Shear transformation:~Cartesian Coordinates}
To find the shear elasticity of the monolayer film, we do a simple shear transformation of the Laplacian by applying the Laplace-Beltrami operator to eq.~\ref{apeq:3D_diffusion}
\begin{equation}
    \Delta_{lb}=\frac{1}{\sqrt{\vert g_{ij}\vert}} \sum_{i j} {\partial_i}\left(g^{i j} \sqrt{\vert g_{ij}\vert} {\partial_j}\right)
\end{equation}
where $g_{ij}$ is the metric tensor and $g^{ij}$ is its inverse. 
The shear transformation to the orthogonal coordinate $(x^\prime,y^\prime)$ in matrix form is 
\begin{equation}
 \begin{bmatrix}x^{\prime}\\y^{\prime}\end{bmatrix}=
 \begin{bmatrix}1&k\\0&\sec\phi\end{bmatrix}\begin{bmatrix}x\\y\end{bmatrix}
\end{equation}
where $k=\tan(\phi)$ and $\phi$ is the shear angle in x-direction. The metric tensor is thus given by
\begin{equation}
    g_{ij}=J^TJ=\begin{bmatrix}1&0\\-k\cos\phi&\cos\phi\end{bmatrix}\begin{bmatrix}1&-k\cos\phi\\0&\cos\phi\end{bmatrix}=\begin{bmatrix}1&-\sin\phi\\-\sin\phi&1\end{bmatrix}.
\end{equation}
Here the determinant of the Jacobian and the metric tensor are $\vert J\vert=\cos\phi$ and $\vert g_{ij}\vert=\cos^2\phi$ respectively, and the inverse is given by
\begin{equation}
    g^{ij}=\frac{1}{\cos^2\phi}\begin{bmatrix}1&\sin\phi\\\sin\phi&1\end{bmatrix}.
\end{equation}
The transformed Laplacian is thus 
\begin{equation}
    \frac{\partial q(\mathbf{r}, s)}{\partial s}=\left(\underbrace{\frac{1}{\cos^2\phi}\frac{\partial^2}{\partial {x^\prime}^2}}_{F_1}+\underbrace{\frac{1}{\cos^2\phi}\frac{\partial^2}{\partial {y^\prime}^2}}_{F_2}+\underbrace{\frac{\partial^2}{\partial {z^\prime}^2}}_{F_3}+\underbrace{\frac{2\sin\phi}{\cos^2\phi}\frac{\partial^2}{\partial x^\prime\partial y^\prime}}_{F_0}\right) q(\mathbf{r}, s)-\underbrace{\omega(\mathbf{r}, s) q(\mathbf{r}, s)}_{F^*}\\
    \label{apeq:3D_diffusion_sheared}
\end{equation}
\subsubsection{Splitting scheme}
In the presence of the mixed derivative term ${F_o}$ in eq.~\ref{apeq:3D_diffusion_sheared}, we use the splitting scheme proposed by Hundsdorfer~\cite{hundsdorferAccuracyStabilitySplitting2002,inthoutStabilityADISchemes2007b}, which solves $Y_o$ explicitly and the rest implicitly.\\
\begin{equation}
    \begin{cases}
Y_0=q_{n-1}+\Delta s F\left( q_{n-1}\right), \\
Y_j=Y_{j-1}+\frac{1}{2}\Delta s\left(F_j\left(Y_j\right)-F_j\left(q_{n-1}\right)\right)-\frac{1}{6}\Delta s\left(F^*(Y_j)-F^*(q_{n-1})\right), &j=1,2,3, \\
\widetilde{Y}_0=Y_0+\frac{1}{2} \Delta s\left(F\left( Y_3\right)-F\left( q_{n-1}\right)\right), \\
\tilde{Y}_j=\widetilde{Y}_{j-1}+\frac{1}{2} \Delta s\left(F_j\left(\widetilde{Y}_j\right)-F_j\left(Y_{3}\right)\right)-\frac{1}{6}\Delta s\left(F^*(\widetilde{Y}_j)-F^*(Y_{3})\right), & j=1,2,3 \\
q_n=\widetilde{Y}_3 
    \end{cases}       
\end{equation}

The fourth-order approximation of the mixed derivative term $F_o$ is simply
\begin{align}
\frac{\partial^2q_{i,j}}{\partial x\partial y}&\approx \frac{\left(1+\beta\right)\left(q_{i+1, j+1}+q_{i-1, j-1}\right)-\left(1-\beta\right)\left(q_{i-1, j+1}+q_{i+1,j-1}\right)}{4 \Delta x \Delta y}\\
     &+\frac{4 \beta q_{i, j}-2 \beta\left(q_{i+1, j}+q_{i, j+1}+q_{i-1, j}+q_{i, j-1}\right)}{4 \Delta x \Delta y}
\end{align}
where $\beta$ denotes a real parameter with $-1 \leqslant \beta \leqslant 1$. The right-hand side is the most general form of a second-order approximation for the cross derivative based on a centered 9-point stencil. When $\beta=0$ reduces to the standard 4-point stencil
\begin{equation}
    \left(q_{x y}\right)_{i, j} \approx \delta_x \delta_y q_{i, j}=\frac{q_{i+1, j+1}+q_{i-1, j-1}-q_{i-1, j+1}-q_{i+1, j-1}}{4 \Delta x \Delta y}
\end{equation}
\end{suppinfo}

\bibliography{refs}

\end{document}